\documentstyle[eqsecnum,aps,twocolumn,psfig,myabstract]{revtex}

\begin{document}
\draft

\long\def\jumpover#1{{}}

\def\approxgt{\,\raise2pt \hbox{$>$}\kern-8pt\lower2.pt\hbox{$\sim$}\,}
\def\approxlt{\,\raise2pt \hbox{$<$}\kern-8pt\lower2.pt\hbox{$\sim$}\,}
\def \qd2{\quad\quad}
\def \th{\thinspace}
\def \ngth{\negthinspace}
\def\ngq{\quad}
\def \ngt2{\negthinspace\negthinspace}
\def \ni{\noindent}
\def \Teff{{$T_{ef\!f} $}}
\def \Mo{{$M_\odot $}}
\def \Lo{{$L_\odot $}}
\def \at{{\rm\char'100}}
\def \dotd{\hbox{$.\!\!^{\rm d}$}}
\def \dotm{\hbox{$.\!\!^{\rm m}$}}

\title{{\rm ASTROPHYSICAL JOURNAL LETTERS, in press, accepted May 18, 1998}
\break
\ \ \break
Nonlinear Beat Cepheid Models}
\author{Zolt\'an Koll\'ath}
\address {Konkoly Observatory, Budapest, HUNGARY}
\author{Jean-Philippe Beaulieu}
\address {Kapteyn Institute, Groningen, The NETHERLANDS}
\author{J. Robert Buchler \th\th \& \th\th Phil Yecko}
\address{Physics Department, University of Florida, Gainesville, FL, USA}


 \address{\ }
 \myabstract{Abstract:\quad 
 The numerical hydrodynamic modelling of beat Cepheid behavior has been
a longstanding quest in which purely radiative models have failed
miserably.  We find that beat pulsations occur naturally when {\it
turbulent convection} is accounted for in our hydrodynamics codes.
 The development of a relaxation code and of a Floquet stability
analysis greatly facilitates the search for and analysis of beat
Cepheid models.
 The conditions for the occurrence of beat behavior can be understood
easily and at a fundamental level with the help of amplitude equations.
Here a discriminant ${\cal D}$ arises whose sign decides whether single
mode or double mode pulsations can occur in a model, and this ${\cal D}$
depends {\sl only} on the values of the nonlinear coupling coefficients
between the fundamental and the first overtone modes.  For radiative
models ${\cal D}$ is always found to be negative, but with sufficiently
strong turbulent convection its sign reverses.
 }





 \keywords{turbulence, convection, hydrodynamics,\hfill\break
 (stars: variables:) Cepheids, stars: oscillations,\hfill\break
 (galaxies:) Magellanic Clouds
 }

 \maketitle

 The Fourier analysis of the observational data of the {\sl beat
Cepheid} light curves and radial velocities shows constant power in two
basic frequencies and in their linear combinations which indicates that
the stars pulsate in two modes (or more if resonances are involved).
Since the beginning of theoretical Cepheid studies in the early 1960s
numerical hydrodynamical attempts at modelling the phenomenon of beat
pulsation have failed, and beat Cepheids have been a bane of stellar
pulsation theory.

 In Cepheids energy is carried through the pulsating envelope to the
surface by radiation transport as well as by turbulent convection (TC).
Even though convection can transport almost all the energy in the
hydrogen partial ionization region, this convection is inefficient in
the sense that it only mildly affects the structure of the envelope.  It
was thus generally thought that convection, while important for
providing a red edge to the instability strip, would play a minor role
the appearance of the nonlinear pulsation.  Purely radiative models did
indeed give good overall agreement with the observed light and radial
velocities.  However, recently it has become increasingly clear that
there are a number of severe problems with radiative models (Buchler
1998), in addition to their inability to account for beat behavior.

 We have recently implemented in our hydrodynamics codes a one
dimensional model diffusion equation for turbulent energy (Yecko,
Koll\'ath \& Buchler 1998) similar to those advocated by Stellingwerf
(1982), Kuhfuss (1986), Gehmeyr \& Winkler (1992) and Bono \&
Stellingwerf (1994).  In contrast to these authors, however, we have
developed additional tools that allow us to find beat behavior without
having to rely on very time-consuming and sometimes inconclusive
hydrodynamic integrations to determine if a model undergoes stable, or
steady beat pulsations.  These are (a) a linear stability analysis which
yields the frequencies and growth rates of {\sl all modes}, (b) a
relaxation method (based on the general algorithm of Stellingwerf with
the modifications of Kov\'acs \& Buchler, 1987) to obtain nonlinear
periodic pulsations (limit cycles) when they exist, (c) a stability
analysis of the limit cycles that gives their (Floquet) stability
exponents.

The 1D turbulent diffusion equation, and the concomitant eddy viscosity,
the turbulent pressure and the convective and turbulent fluxes contain
(seven) order unity parameters that need to be calibrated through a
comparison to observations.  In a first paper (Yecko, Koll\'ath \&
Buchler 1998) in which we performed a broad survey of the linear
properties of TC Cepheid models we found that of these the mixing
length, the strengths of the convective flux and of the eddy viscosity
play a dominant role and that broad combinations of these three
parameters exist that give agreement with the observed widths of both
the fundamental and first overtone instability strips.

 In this Letter we show that the inclusion of TC produces pulsating beat
Cepheid models that satisfy the observational constraints, in particular
those of period ratios, of modal pulsation amplitudes and of their
ratios.  Furthermore the models are very robust with respect to the
numerical and physical parameters.


Our discovery of beat Cepheid models has been partially serendipitous.
When we started to investigate the nonlinear pulsations of a typical
Small Magellanic Cloud Cepheid model ($M\ngth =\ngth 4.0M_\odot$,
$L\ngth =\ngth 1100L_\odot$, \Teff = 5900\th K, $X\ngth =\ngth 0.73$ and
$Z\ngth =\ngth 0.004$) with the the turbulent convective hydrocode, we
encountered beat pulsations that appeared steady.  We use the OPAL
opacities of Iglesias \& Rogers (1996) combined with those of Alexander
\& Ferguson (1994).  The values of the TC parameters -- for a definition
cf. Yecko et al. (1998) -- are $\alpha_c\ngth =\ngth 3$, $\alpha_\Lambda
\ngth =\ngth0.41$, $\alpha_p\ngth =\ngth 0.667$, $\alpha_t\ngth =\ngth
1$, $\alpha_D\ngth =\ngth 4$, $\alpha_s\ngth =\ngth 0.75$, $\alpha_\nu
\ngth =\ngth 1.2$.  The steadiness of these beat pulsations was
confirmed when several nonlinear hydrodynamics calculations, each
initiated with a different admixture of fundamental and first overtone
eigenvectors, converged towards the same final steady beat pulsational
state.  This convergence could be corroborated when we extracted the
slowly varying amplitudes with the help of a time-dependent Fourier
decomposition, and plotted the resulting phase portraits ($A_0(t)$
vs. $A_1(t)$) that are shown in Fig.~1 where all initializations are
seen to converge toward a fixed point located at $A_0 \ngth =\ngth
0.0104$ and $A_1\ngth =\ngth 0.0200$ (These radial displacement
amplitudes assume the eigenvectors to be normalized to unity at the
stellar surface, $\delta r/r_*\ngth =\ngth 1$).

\vspace{10pt}

 \psfig{figure=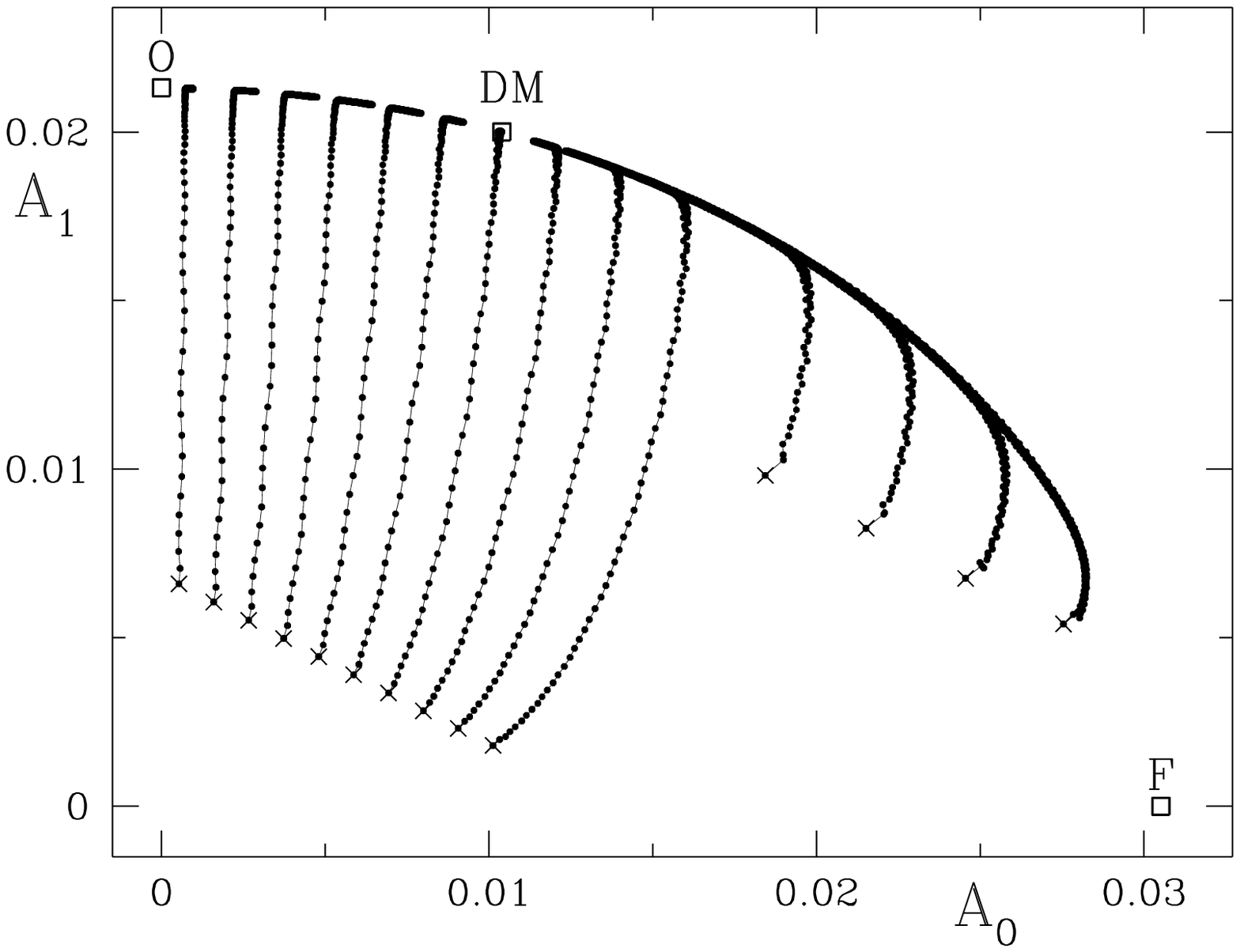,width=8.5cm}
 \vspace{5pt}
{\parindent=0pt
\parbox{8.7cm}{\small\baselineskip=5pt
 \noindent{\bf Fig.1:\ } 
 Evolution of the modal amplitudes for different initial conditions
(marked with a cross).  Equal time intervals between dots.  The open
squares denote the unstable fundamental (F), the unstable first overtone
(O) and the stable double-mode (DM) fixed points.
 }}
 \vspace{10pt}

 While the observed transient behavior of the models provides a
conclusive proof of the presence of steady beat pulsations, it is
important to explain and describe the behavior on a more fundamental
level.  The phase portrait of Fig.~1 is very similar to those found for
nonresonant mode interaction on the basis of amplitude equations.
(Buchler \& Kov\'acs 1986, 1987, hereafter BK86 and BK87).
 We show here that indeed the nonlinear behavior of the hydrodynamical
model pulsations can be understood very simply that way.

The amplitudes of the two nonresonantly interacting modes obey
remarkably simple equations
 \begin{eqnarray}
 \quad\quad {dA_0 \over dt} = & A_0 \th (\kappa_0 - q_{00} A_0^2 
  - q_{01} A_1^2) \nonumber \quad\quad\quad\quad\quad (1a)\\
 \quad\quad  {dA_1\over dt}  = & A_1 \th (\kappa_1 - q_{10} A_0^2
  - q_{11} A_1^2) \nonumber \quad\quad\quad\quad\quad (1b)
 \end{eqnarray}
 These amplitude equations are `normal forms' and are therefore generic
for any dynamical system in which two modes interact nonresonantly.  The
assumptions underlying these amplitude equations are satisfied for
Cepheids:
 (a) The lowest modes (fundamental and first overtone here) are weakly
nonadiabatic, i.e. the ratios of linear growth rates $\kappa$ to
periods are small, a condition that is readily confirmed by our linear
stability analysis;
 (b) The pulsations are weakly nonlinear which allows a truncation of
the amplitude equations in the lowest permissible (third) order; weak
nonlinearity can be established by comparing the linear and nonlinear
periods which differ less than a tenth of a percent.  Furthermore both
the nonlinear self-saturation coefficients $q_{00}$ and $q_{11}$ as well
as the cross-coupling coefficients $q_{01}$ and $q_{10}$ have always
been found to be positive in Cepheid models so that amplitude saturation
can occur in third order, and it is sufficient to keep terms up to cubic
in the amplitudes.
 (c) In the range of interest there is no important low order resonance
between the fundamental and the first overtone modes, and possibly a
higher mode.

In the following discussion we look at the regime where both modes are
linearly unstable, $\kappa_0>0$ and $\kappa_1> 0$.  Eqs.~(1) then have
two {\sl single mode} fixed points.  The amplitude of the single mode
fundamental (0) fixed point is $A_0=\sqrt{\kappa_0/q_{00}}$ and its
coefficient for stability to first overtone perturbations is
 $\bar\kappa_{1(0)} = \kappa_1 - q_{10} A_0^2 \th$.  In our notation a
positive coefficient implies growth and thus instability.  The
corresponding first overtone (1) limit cycle amplitude is
$A_1=\sqrt{\kappa_1/q_{11}}$ and its coefficient of stability to
fundamental perturbations is $\bar\kappa_{0(1)} = \kappa_0 - q_{01}
A_1^2 \th$.  The $\bar\kappa$'s, when multiplied by the periods $P_k$ of
their limit cycles, are equal to the corresponding Floquet exponents
(Buchler, Moskalik \& Kov\'acs 1991).

 Eqs.~(1) can also have {\sl a double mode fixed point} whose amplitudes
satisfy
  $A_{0\th DM}^2 = \bar\kappa_{0(1)} q_{11}/ {\cal D} < A_0^2$,
  $A_{1\th DM}^2 = \bar\kappa_{1(0)} q_{00}/ {\cal D} < A_1^2$,
 where ${\cal D} = q_{00} q_{11}$ -- $q_{01} q_{10}$.  This fixed point
exists provided $A_{0\th DM}^2>0$ and $A_{1\th DM}^2>0$.
 Then, if ${\cal D} <$ 0, the double mode limit cycle is unstable.
Stable pulsations occur either in the fundamental {\it or} first
overtone, and the pulsational mode is determined by the evolutionary
history of the model (hysteresis).
 On the other hand, if ${\cal D}>0$, the double mode fixed point is
stable, and steady double mode pulsations occur (BK86).  One can show
that these conditions are equivalent to requiring $\bar\kappa_{0(1)}>0$
and $\bar\kappa_{1(0)}>0$, conditions which also imply that both {\sl
single mode} limit cycles (fundamental and first overtone) are
individually unstable.  This validates Stellingwerf's (1975) suggestion
that the simultaneous instability of the fundamental and the first
overtone leads to steady beat pulsations (in the absence of a
resonance).  It provides an economical tool to search for double mode
behavior, because we can now relatively easily compute single mode limit
cycles and their stability.

As a further confirmation that the nonresonant scenario applies to the
pulsating Cepheid model, we have determined the coefficients of Eqs.~(1)
as in BK87 by fitting time-dependent solutions of these equations to the
temporal variation of the amplitudes in their approach to the limit
cycle as shown in Fig.~1.  The fitted trajectories in the phase portrait
are practically undistinguishable from the hydro results, confirming the
applicability and accuracy of the amplitude equation formalism and the
absence of any relevant resonances.

We mention in passing that the expression `double mode Cepheids' is
often used cavalierly for beat Cepheids.  Since no additional, resonant
overtone is involved in the beat pulsations, the latter are thus truly
double mode pulsations.

With the relaxation code we are able to compute both the fundamental and
the first overtone limit cycles with their respective amplitudes and
Floquet stability exponents $\lambda_{1(0)} = P_0 \bar\kappa_{1(0)}$ and
$\lambda_{0(1)} = P_1 \bar\kappa_{0(1)}$.  The above discussion then
shows that from these four quantities we can extract the four nonlinear
$q_{jk}$ coefficients when we have already computed the linear periods
and growth rates.  The values we obtain this way for this beat Cepheid
model agree quite well with those that we obtain from the fit described
in the previous paragraph.  Note that these two determinations rely on
independent numerical hydrodynamical input, the first on two periodic
limit cycles (that are linearly unstable), the second on transient
evolution toward the stable double mode pulsation.

\vskip 10pt

In order to investigate the robustness of the observed beat behavior we
now explore the pulsational behavior of a {\sl sequence} of Cepheid
models in which the effective temperature of the equilibrium modes of
the sequence varies from \Teff\th =\th 6200\th K to 5800\th K.  Such a
sequence is approximately along an evolutionary path.  The eddy
viscosity parameter $\alpha_\nu$ is treated as an additional variable
parameter to explore the effect of TC on the behavior.

\vspace{20pt}
 \psfig{figure=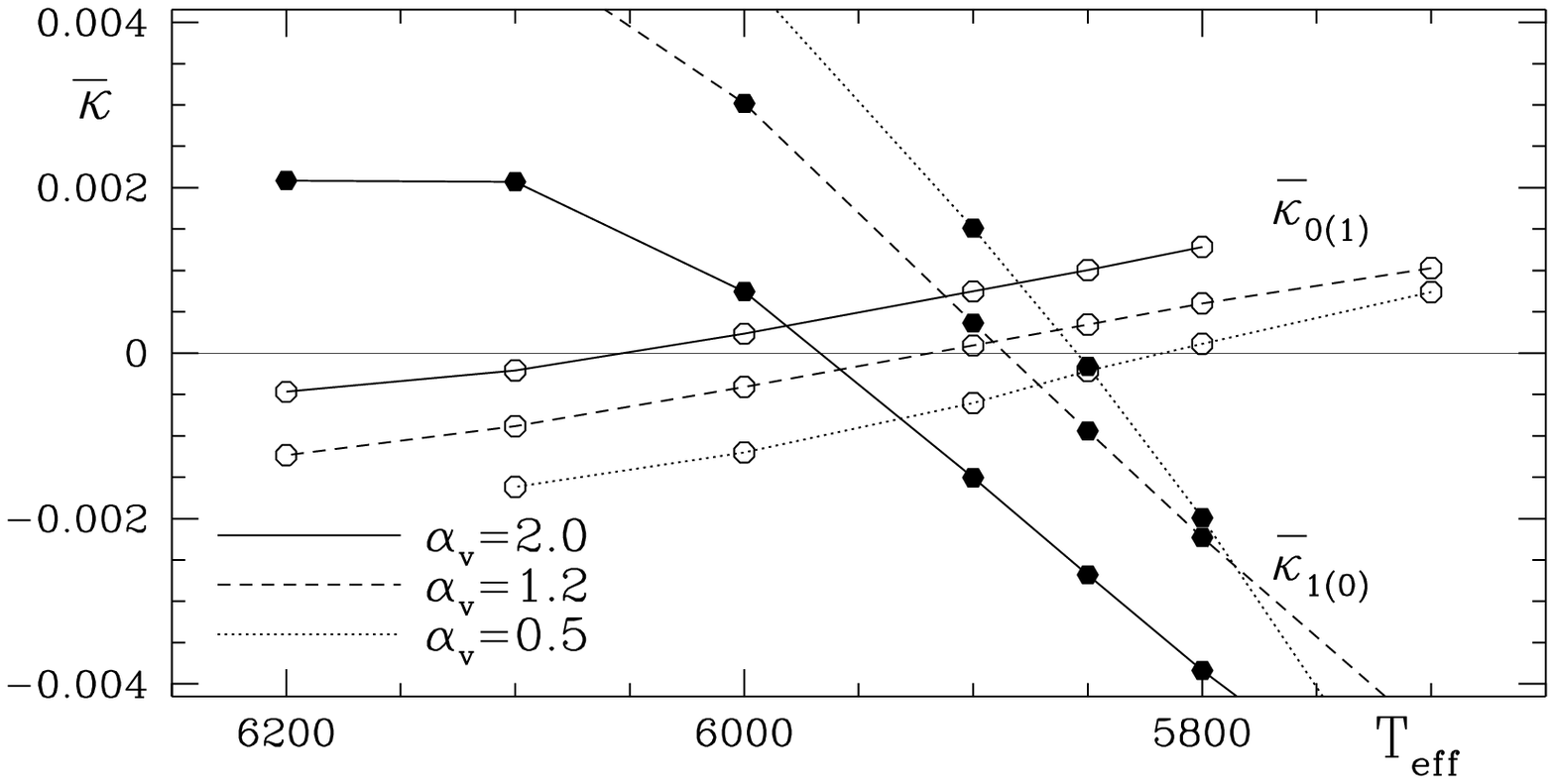,width=8.7cm}
 {\parindent=0pt
\parbox{8.7cm}{\small\baselineskip=5pt
 \noindent {\bf Fig.2:\ }
 The limit cycle stability coefficients ($d^{-1}$) along a
Cepheid sequence (i.e. as a function of \Teff) for various eddy
viscosity strengths $\alpha_\nu$; \th fundamental limit cycles
$\bar\kappa_{1(0)}$ (full circles), first overtone limit cycles
$\bar\kappa_{0(1)}$ (open circles).
 }}
\vspace{15pt}

In Fig.~2 the stability coefficients of the sequence are plotted versus
\Teff, with open/filled circles for those of the fundamental/overtone
single mode cycles.  The curves are labelled with the corresponding
strengths $\alpha_\nu$ of the eddy viscosity.  As discussed above we
expect double mode behavior where both Floquet exponents are
positive. (The stability exponents due to perturbations with other modes
are always smaller in this sequence and are therefore irrelevant here).
For the low value of $\alpha_\nu \ngth =\ngth 0.5$ (dotted lines) the
two stability coefficients are never positive simultaneously, thus
excluding double-mode behavior.  On the other hand, for $\alpha_\nu =
1.2$ a double mode region appears between \Teff $\sim$ 5875 -- 5915\th K
and for $\alpha_\nu \ngth =\ngth 2.0$ this broadens to \Teff $\sim$ 5965
-- 6050\th K.


How does turbulent convection bring about double mode behavior?  Fig.~2
shows that, in the region of interest, an increase in the turbulent eddy
viscosity causes a rapid decrease in the stability of the fundamental
limit cycle ($\bar\kappa_{1(0)}$, filled circles), but an increase in
that of the first overtone limit cycle ($\bar\kappa_{0(1)}$, open
circles).  This description, though, does not tell us whether it is the
effect of TC on the linear $\kappa$'s or on the nonlinear $q$'s, or on
both, that is responsible for the beat pulsations.

Table~1 shows the variation with $\alpha_\nu$ of the relevant model
quantities, viz. the linear growth rates, the nonlinear coupling
coefficients, the discriminant ${\cal D}= q_{00} q_{11} - q_{01}q_{10}$,
the amplitude of the fundamental cycle and its stability coefficient
with respect to overtone perturbations, and the same for the first
overtone.

The {\sl necessary} condition for stable double mode pulsations, viz.
 ${\cal D} >$ 0, is never found to be satisfied in radiative models.  In
these models the cross-coupling always dominates over the
self-saturation coefficients.  Table~1 shows that an increase in the
strength of the eddy viscosity causes $q_{00}$ and $q_{11}$ to increase
faster than $q_{01}$ and $q_{10}$.  It is therefore the resultant change
in the sign of ${\cal D}$ that makes double mode behavior possible for
sufficiently large $\alpha_\nu$.

The condition for beat behavior is thus seen to be rather subtle in that
it involves the effects of convection beyond the linear regime for which
it seems difficult to give a `simple' physical explanation.

Fig.~3 gives the overall modal selection picture in the $\alpha_\nu$ --
\Teff\ plane.  The linear edges of the instability region
($\kappa_0\ngth =\ngth 0$ and $\kappa_1\ngth =\ngth 0$) are shown as
dashed lines.  By computing the fundamental and first overtone limit
cycles for a number of $\alpha_\nu$ and \Teff\ values, by interpolation,
we can obtain $\bar\kappa_{0(1)}$ or $\bar\kappa_{1(0)}$ as a function
of $\alpha_\nu$ and \Teff, and in particular the loci where they vanish.
The solid curves give the nonlinear pulsation edges and are marked ORE
and FBE.

 It is straightforward to show that if the two linear growth rates
vanish at the same point, the four curves will intersect in a single
point on this diagram, that we label {\sl critical point}.

The curve marked OBE is the linear blue edge of the first overtone mode
and it coincides with the overtone nonlinear blue edge up to and on the
left of the critical point.  The linear fundamental blue edge becomes
also the fundamental blue edge above the critical point.  Above the line
ORE we have $\bar\kappa_{0(1)} \ngth > \ngth 0$ and the first overtone
limit cycle is unstable.  Below the line FBE the quantity
$\bar\kappa_{1(0)}\ngth > \ngth 0$ and the fundamental limit cycle is
unstable.  Thus in the region marked `dm' both single mode limit cycles
are unstable, and this is the region of double-mode pulsation.  In the
small triangular region at the bottom, on the other hand, both limit
cycles are stable, and {\sl either} fundamental or first overtone limit
cycles can occur.

\vspace{15pt}
 \psfig{figure=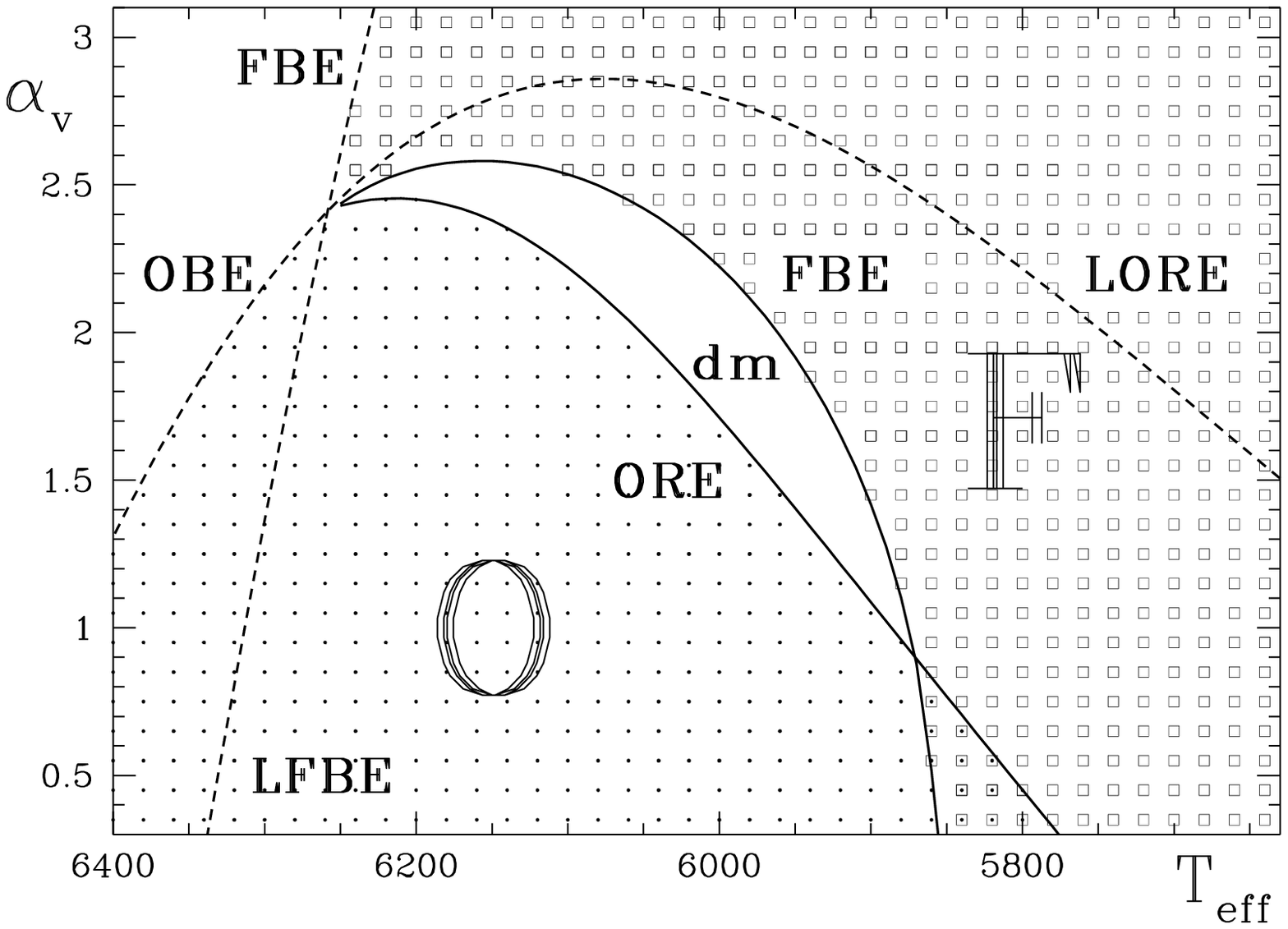,width=8.7cm}
 {\parindent=0pt
 \parbox{8.7cm}{\small\baselineskip=5pt
 \noindent {\bf Fig.3:\ }  
 Modal selection in the $\alpha_\nu$ -- \Teff\ plane.
 }}
\vspace{5pt}

In summary, stable first overtone pulsations occur in the dotted region,
delineated by the lines OBE and ORE.  The fundamental limit cycle is
stable in the region marked by open squares, delineated by FBE and FRE
(not shown on the far right).  This figure makes it particularly evident
how TC favors double mode pulsations and why all efforts with radiative
codes have failed in modelling beat Cepheids.

\vskip 10pt

We have seen that when TC effects are sufficiently large then the
Cepheids should run into the double mode regime in both their crossings
of the instability strip.  Furthermore, as a Cepheid crosses the double
mode regime redward, say, the first overtone amplitude should gradually
go to zero while the fundamental amplitude increases from zero to the
value it attains as a fundamental mode Cepheid (BK86).  The question
arises whether this nonresonant scenario that is derived from the
amplitude equations is in agreement with the observations.

The four SMC beat Cepheids from the EROS survey (analyzed by Beaulieu
and reproduced in Buchler 1998) all have the same amplitude ratios,
$A_0/A_1 \sim 0.45$, a priori in disagreement with the nonresonant
scenario shown in Fig.~1. of BK86 that suggests that Cepheids with all
amplitudes ratios should occur. 

 In Fig.~4 we display the behavior of the component modal amplitudes of
the beat Cepheid models for the $\alpha_\nu\ngth =\ngth 1.2$ sequence of
Fig.~2.  The amplitudes of the stable single mode limit cycles are shown
as solid lines with solid dots for the fundamental and open dots for the
first overtone, and as dashed lines where they are unstable.  The
fundamental and first overtone component amplitudes of the stable double
mode pulsators are shown as solid squares and open diamonds,
respectively.  It is seen that although the modal amplitudes do indeed
vary continuously throughout as the double mode regime is traversed, the
behavior is very rapid near the cooler side.  The reason for this
unexpected behavior is that the q's are {\sl not} constant in this
sequence, and what is more, they vary in such a way that ${\cal D}$
happens go through zero around 5850\th K.  It is the presence of this
nearby pole that causes a change in the curvature of $A_0$.

\vspace{15pt}
 \psfig{figure=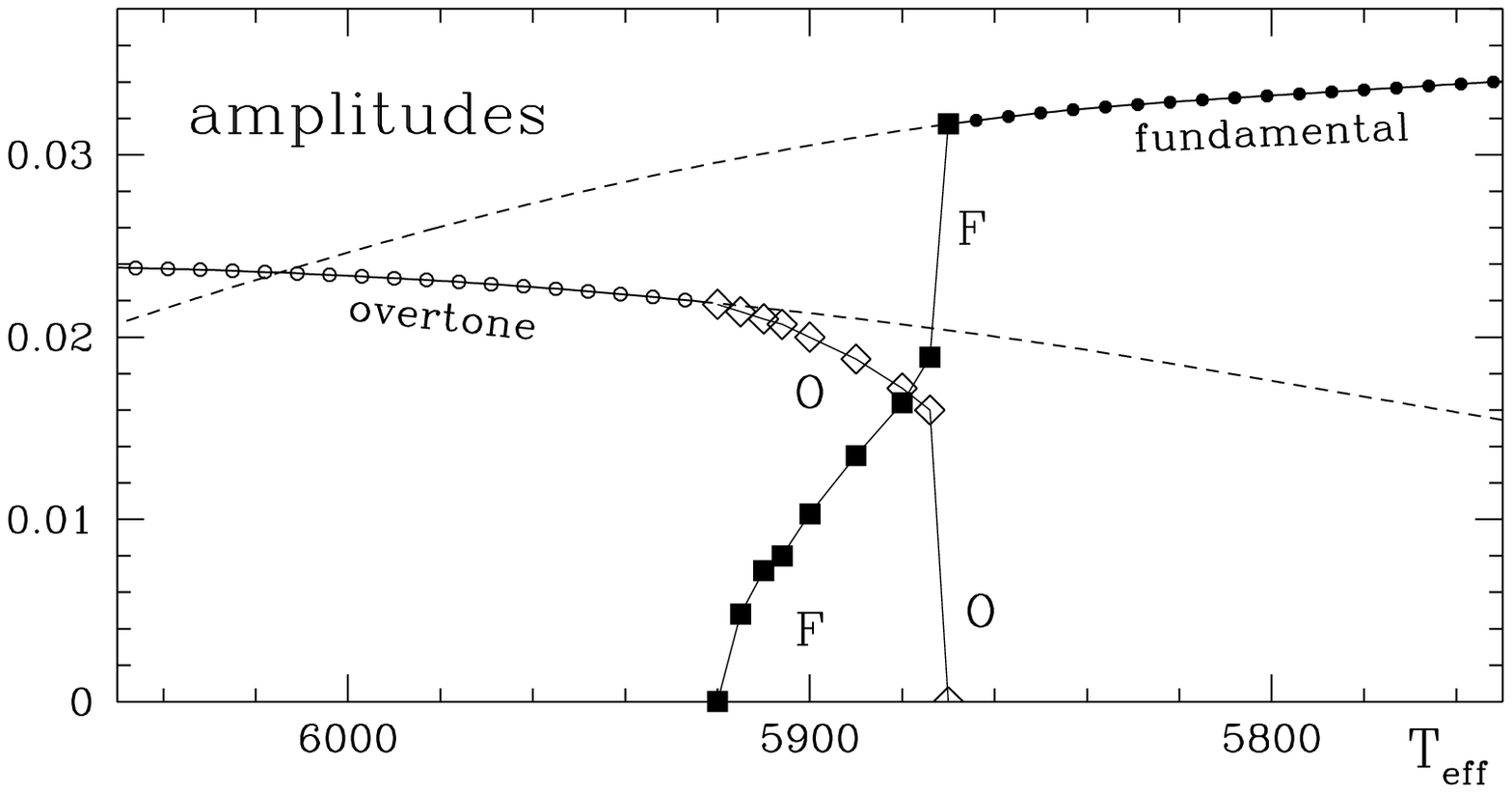,width=8.7cm}
\vspace{10pt}
 {\parindent=0pt
 \parbox{8.7cm}{\small\baselineskip=5pt
 \noindent {\bf Fig.4:\ }  
 Sequence of models (same as in top of Fig.~2). Amplitudes of
the fundamental and of the first overtone limit cycles (lines),
solid/dashed where stable/unstable.
 Double mode component amplitudes of the double mode (fundamental: solid
squares and first overtone: open diamonds).
 }}
\vspace{15pt}

According to Fig.~4 it is therefore much more likely to find beat
Cepheids in the slowly varying regime where the ratio $A_0/A_1\approxlt
0.5$.  The computed behavior of the modal amplitudes is thus in
agreement with the observed SMC Cepheids, and the nonresonant scenario
is consistent with the observations.

\vskip 10pt

We have demonstrated that turbulent convection leads naturally to beat
behavior in Cepheids, which does not occur with purely radiative models.
The reason is that the nonlinear effects of TC dissipation can create a
region in which both the fundamental and the first overtone cycles are
unstable, and the model undergoes stable double mode pulsations.  At a
more basic level the amplitude equation formalism shows that turbulent
convection modifies the nonlinear coupling between the fundamental and
first overtone modes in such as way as to allow beat behavior.

The development of a relaxation code (TC) to find periodic pulsations,
and a Floquet stability analysis of these limit cycles has made the
search quite efficient, and a broader survey of beat Cepheids, with wide
ranges of metallicities is in progress.  This will also search for beat
Cepheid models that pulsate in the first and second overtones.

\vskip 10pt

{\small 
 This research has been supported by the Hungarian OTKA (T-026031), AKP
(96/2-412 2,1) and by the NSF (AST95--28338) at UF.  Two of us (JPB) and
(ZK) thank the French Acad\'emie des Sciences for financial support.  }

\clearpage

\widetext
 \vspace{13pt}
 {\noindent TABLE 1.\-- \  Nonlinear Coupling Coefficients
(5900\th K model).}

\vspace{10pt}

{\small
 \begin{tabular}{rrrrrrrrrrrr}
 \hline\hline
 \hfill\ $\alpha_\nu$ \hfill
 &\hfill $\kappa_0$\ \ \ \hfill &\hfill $\kappa_1$\ \ \ \hfill
  &\hfill  $q_{00}$ \hfill &\hfill $q_{01}$ \hfill &\hfill $q_{10}$
\hfill &\hfill  $q_{11}$ \ \hfill &\hfill ${\cal D}$ \hfill 
&\hfill  $A_0$ \ \ \hfill &\hfill $\bar\kappa_{0(1)}$\ \  \hfill
 &\hfill  $A_1$ \ \ \hfill &\hfill $\bar\kappa_{1(0)}$\ \ \hfill \\
 \hline \noalign{\smallskip\smallskip\smallskip}
 0.5 &\ngq 2.382e-3 &\ngq  1.124e-2 &\ngq  1.199 &\ngq  3.532 &\ngq 
 4.898 &\ngq  13.311 &\ngq --1.343 &\ngq
 4.458e-2 &\ngq 1.510e-3 &\ngq 2.906e-2  &\ngq  --6.012e-4   \\
 1.0 & \ngq 2.169e-3 & \ngq 8.636e-3 & \ngq 1.815 & \ngq 3.865 &
\ngq 6.645 & \ngq 14.924 & \ngq 1.407 & \ngq
 3.457e-2 & \ngq 6.959e-4 & \ngq 2.406e-2 & \ngq --6.752e-5 \\
 1.2 &\ngq 2.082e-3 &\ngq  7.582e-3 &\ngq 2.179 &\ngq  4.200 &\ngq 
 7.554 &\ngq  16.001 &\ngq  3.140 &\ngq 
 3.091e-2 &\ngq 3.650e-4&\ngq 2.177e-2 &\ngq  9.169e-5   \\
 1.5 &\ngq 1.947e-3 &\ngq 5.983e-3 &\ngq 2.893 &\ngq 4.917 &
\ngq 9.212 & \ngq 18.172 & \ngq 7.269 & \ngq
 2.595e-2 & \ngq --2.188e-4 & \ngq 1.814e-2 & \ngq 3.285e-4  \\
 2.0 &\ngq  1.720e-3 &\ngq 3.282e-3 &\ngq  4.685 &\ngq  7.194 &\ngq
 13.039 &\ngq  24.229 &\ngq  19.704 &\ngq 
 1.916e-2 &\ngq --1.506e-3 &\ngq 1.164e-2 & \ngq 7.458e-4  \\
 \noalign{\smallskip}
 \hline
 \hline
 \end{tabular}
 }
\narrowtext




\begin{references}

\vspace{-1.6cm}


\small

\vskip 2pt \bibitem[]{}
 Alexander, D. R., Ferguson, J. W. 1994,
  ApJ 437, 879
 
\vskip 2pt \bibitem[]{}
 Bono, G., Stellingwerf, R.F. 1994,
    ApJ Suppl 93, 233--269

\vskip 2pt \bibitem[]{}
 Buchler, J.R. 1998, in {\sl A Half Century of Stellar Pulsation
Interpretations: A Tribute to Arthur N. Cox}, ed. P.A. Bradley and
J.A. Guzik, ASP Conf. Ser. 135, 220

\vskip 2pt \bibitem[]{}
    Buchler, J.R., Koll\'ath, Z., Beaulieu, J.P. , Goupil. M.J. 1996,
    ApJ Letters, 462, L83

\vskip 2pt \bibitem[]{}
  Buchler, J.R., , Kov\'acs, G. 1986, ApJ 308, 661, [BK86]; 1987, ibid.
318, 232 [BK87]

\vskip 2pt \bibitem[]{}
 Buchler, J.R., Moskalik, P., Kov\'acs, G. 1991, ApJ 380, 185.

\vskip 2pt \bibitem[]{}
    Gehmeyr, M. , Winkler, K.-H. A. 1992,
    AA 253, 92--100; ibid. 253, 101--112

\vskip 2pt \bibitem[]{}
 Iglesias, C. A.\& Rogers, F. J. 1996,
  ApJ 464, 943
 
\vskip 2pt \bibitem[]{}
 Kov\'acs, G. , Buchler, J.R. 1987, ApJ 324, 1026.
 
\vskip 2pt \bibitem[]{}
   Kuhfuss, R. 1986,
   AA 160, 116

\vskip 2pt \bibitem[]{}
    Stellingwerf, R.F. 1975,
    ApJ 199, 705

\vskip 2pt \bibitem[]{}
    Stellingwerf, R.F. 1982,
    ApJ 262, 330

\vskip 2pt \bibitem[]{}
 Yecko, P., Koll\'ath Z., Buchler, J. R. 1998, A\&A, in press




 \end{references}
\end{document}